# A Note on the Definition of Strategies in Opportunity Hunting Games [*]

Ran Eilat[†]    Zvika Neeman[‡]    Eilon Solan[§]

March 30, 2025

In this note, we propose an alternative definition of strategies and histories in opportunity hunting games, as introduced in Eilat, Neeman and Solan (2025). The advantage of the formulation presented here is that it encompasses a broader class of strategies than the inertial strategies analyzed in Eilat, Neeman and Solan (2025), including those that allow for a countably infinite number of inspections within finite time intervals. Although the definitions of histories and strategies in this broader setting are more subtle, we show that the main results continue to hold. This demonstrates that the findings in Eilat, Neeman and Solan (2025) are robust to environments in which players have access to a richer strategy space.

Specifically, consider the model of Opportunity Hunting Games as presented in Eilat, Neeman and Solan (2025). Suppose we wish to allow a player to inspect at times $\frac{1}{2}, \frac{2}{3}, \frac{3}{4}, \frac{4}{5}, \ldots, 1\frac{1}{2}, 1\frac{2}{3}, 1\frac{3}{4}, 1\frac{4}{5}, \ldots$. This sequence of inspections is not possible under an inertial strategy. While such an inspection policy is presumably suboptimal, we seek definitions of strategies and histories broad enough to accommodate it. The challenge lies in formulating these notions so that they are both general and natural, while also ensuring that any pair of strategies induces a unique distribution over the set of plays. This, in turn, guarantees that the expected payoffs associated with any strategy profile are well defined.

We begin by defining the notion of history.

---

[*]We thank Daniel Bird, Yeon-Koo Che, Laura Doval, Kfir Eliaz, Yingi Guo, Johannes Hörner, John Levy, Raphaël Lévy, Elliot Lipnowski, Qingmin Liu, Ady Pauzner, Alessandro Pavan, Jacopo Perego, Andrea Prat, Ludvig Sinander, Steven Schochet, Jan Zapal, and seminar participants in Bar-Ilan University, Columbia University, Tel-Aviv University, the University of Glasgow, the University of Tokyo, the 10th Israeli IO Workshop, the 24th ACM Conference on Economics and Computation (EC'23), and ESSET 2024, for helpful comments. Eilat and Neeman acknowledge the support of the Israel Science Foundation, Grant #1792/23. Solan acknowledges the support of the Israel Science Foundation, Grant #211/22.

[†]Department of Economics, Ben-Gurion University of the Negev, eilatr@bgu.ac.il
[‡]School of Economics, Tel Aviv University, zvika@tauex.tau.ac.il
[§]The School of Mathematical Sciences, Tel Aviv University, eilons@tauex.tau.ac.il

**Definition (History).** *A history is a 3-tuple $\langle \alpha^*, \{t_\alpha\}_{\alpha \leq \alpha^*}, \iota \rangle$ where: (i) $\alpha^*$ is a countable ordinal;[1] (ii) $\{t_\alpha\}_{\alpha \leq \alpha^*}$ is an increasing sequence of non-negative real numbers, such that $t_\alpha = \lim_{\beta < \alpha} t_\beta$ for every limit ordinal $\alpha \leq \alpha^*$; and (iii) $\iota$ is a function that assigns two nonempty sets of players, $\iota_a$ and a singleton set $\iota_i$, to every successor ordinal $\alpha \leq \alpha^*$.*

The interpretation of a history is as follows. The players' inspection times before $t_{\alpha^*}$ are given by

$$\{t_\alpha : \alpha \leq \alpha^*, \alpha \text{ is a successor ordinal}\}.$$

For each time $t_\alpha$ in this set, the players in $\iota_a(\alpha)$ are the players who attempted an inspection at that time, and the player in the singleton set $\iota_i(\alpha)$ is the player who actually inspected at that time.[2,3]

**Example.** To illustrate the definition above, consider a history in which a player inspects at times $\frac{1}{2}, \frac{2}{3}, \frac{3}{4}, \ldots$, and then at times $1\frac{1}{2}, 1\frac{2}{3}, 1\frac{3}{4}, \ldots$. To describe this history, we set $\alpha^* = 2\omega$. Then, $t_0 = 0, t_1 = \frac{1}{2}, t_2 = \frac{2}{3}, t_3 = \frac{3}{4}, \ldots, t_\omega = 1, t_{\omega+1} = 1\frac{1}{2}, t_{\omega+2} = 1\frac{2}{3}, t_{\omega+3} = 1\frac{3}{4}, \ldots$, and $t_{2\omega} = 2$. In this description, $0, \omega$ and $2\omega$ are limit ordinals. All other ordinals, namely $1, 2, 3, \ldots, \omega + 1, \omega + 2, \omega + 3, \ldots$, are successor ordinals, each associated with a specific inspection time. In this example, since only one player inspects during this history, $\iota_a = \iota_i$ is the singleton set containing this player for every successor ordinal. ∎

Next, we define the notion of strategy. A strategy is a measurable function from histories to distributions over future inspection times. For measurability to be well defined, we endow both of these spaces with appropriate sigma-algebras. The sigma-algebra over the space of distributions on $[0, \infty]$ is the weak-$*$ topology. The sigma-algebra over the space of histories is the one inherited from $\cup_\alpha \mathbb{R}^\alpha$, where the union is over all countable ordinals. Formally,

**Definition (Strategy).** *A strategy $\sigma_i$ of player $i$ is a measurable function that assigns to every history $h = \langle \alpha^*, \{t_\alpha\}_{\alpha \leq \alpha^*}, \iota \rangle$ a probability distribution over $(t_{\alpha^*}, \infty]$. An atom of $\sigma_i(h)$ at $\infty$ is interpreted as if the player assigns positive probability to never inspecting.*

A *play* is an object that indicates how the game was played from beginning to end. In our model, the definition of play coincides with that of history, with one difference: when $\langle \alpha^*, \{t_\alpha\}_{\alpha \leq \alpha^*}, \iota \rangle$ is a history, $\{t_\alpha\}_{\alpha \leq \alpha^*}$ is a collection of non-negative real numbers. In a play, we allow $t_{\alpha^*}$ to be equal to $\infty$. This is because along the play the players may continue inspecting ad infinitum and never discover the prize. We endow the space of plays with the sigma-algebra inherited from $\cup_\alpha (\mathbb{R} \cup \{\infty\})^\alpha$, where the union is over all countable ordinals.

---

[1] Ordinal numbers are linearly ordered labels that include the natural numbers and have the property that every set of ordinals has a least element. This facilitates the definition of an ordinal number $\omega$ that is greater than every natural number, along with ordinal numbers $\omega + 1, \omega + 2, \ldots$, which are greater than $\omega$. Similarly, the ordinal number $2\omega$ is greater than $\omega$ plus every natural number, and so forth.

[2] By definition, no player inspects at times $t_\alpha$ for *limit* ordinals $\alpha$. It is possible to also consider a more general set of strategies where players may also inspect at times that are associated with limit ordinals. Doing so would not affect our results.

[3] The requirement that $\alpha^*$ is countable is without loss of generality, because between each $t_\alpha$ and $t_{\alpha+1}$ there is at least one rational number, and the set of rational numbers is countable.

The next result ensures that every pair of strategies induces a well defined distribution over plays. This implies that the players' expected payoffs from any pair of strategies are well defined.

**Lemma 1.** *Every pair of strategies $(\sigma_1, \sigma_2)$ induces a unique probability distribution over plays.*

Consider the definition of a Markov strategy as given in Eilat, Neeman and Solan (2025). The following result shows that, to verify whether a given pair of Markov strategies constitutes a Markov Perfect Equilibrium (MPE), it is sufficient to check that no player can profitably deviate to another Markov strategy. While this property is well known in discrete-time models, in our continuous-time setting the argument is more subtle.

**Proposition 1.** *A pair of Markov strategies $(\sigma_1, \sigma_2)$ is an MPE if and only if no player has a profitable deviation to another Markov strategy.*

An immediate consequence of Proposition 1 is that all symmetric MPEs characterized in Eilat, Neeman and Solan (2025) remain equilibria, even when players are allowed to deviate to any strategy in the broader class introduced in this note.

NOTE ON THE LITERATURE. The literature on continuous-time games has proposed various approaches for modeling histories and strategies. In the context of our model, our approach offers several advantages compared to the existing body of work.

For example, in the literature on optimal control (e.g., Cardaliaguet (2007)) a player's strategy specifies a grid of time points. Starting at time zero, and for any time point on the grid, the player "commits to" a plan of action, which may depend on the history of play, to be played until the next time point on the grid. Applying this method in our model would only permit the characterization of $\varepsilon$-equilibria.

Other papers (e.g., Bergin and MacLeod (1993)) have employed the concept of "inertia" in a strategy. Specifically, once an action is played, the player cannot switch to another action within a small interval of time.[4] Papers in this literature study the limit of equilibria as the length of the inertia interval decreases to zero. In contrast, our approach does not rely on the implicit assumption that a player can act only finitely many times in any finite time interval, and does not need to address the issue of whether the equilibrium correspondence is continuous as the length of the inertia interval vanishes.[5]

---

[4]Alternatively, a player cannot repeat the action within the mentioned time interval.

[5]Kamada and Rao (2023) study an environment in which focusing on standard "inertial" strategies may be inadequate. In their framework, the action of each player can depend on the realization of a stochastic process and on the history of play at each point in time. They identify restrictions on strategies and histories that guarantee that each strategy profile induces a unique path of play.

Stinchcombe (1992) proposes a definition for strategy that, similar to ours, employs the concept of ordinals. A key distinction between our definition and Stinchcombe's (1992) is that, in our framework, actions are instantaneous and occur at times associated with each ordinal. In contrast, in Stinchcombe (1992), a player maintains the same action between any two times associated with consecutive ordinals. This distinction makes our definition more useful for games in which the payoff from actions is "discrete" (e.g., a cost of inspection).

Finally, Fudenberg and Tirole (1985) analyze the limit of a sequence of games that are played on a discrete time grid. In those cases where solving the game in continuous time is simpler, as in our case, our approach is more straightforward.

## PROOF OF LEMMA 1

We will construct the unique probability distribution over plays induced by $(\sigma_1, \sigma_2)$ by transfinite induction. $t_0$ is set to 0.

Let $\alpha_*$ be a limit ordinal, and suppose that the distribution over histories with $\alpha < \alpha_*$ is uniquely defined. Then $t_{\alpha_*} = \sup_{\alpha < \alpha_*} t_\alpha$.

Let $\alpha_* + 1$ be a successor ordinal, and suppose that the distribution over histories with $\alpha \leq \alpha_*$ is uniquely defined. Denote by $h$ the random variable of the history up to $t_{\alpha_*}$. The conditional distribution of $t_{\alpha_*+1}$ given $h$ is determined by $\sigma_1(h)$ and $\sigma_2(h)$.

## PROOF OF PROPOSITION 1

We show that for every $\varepsilon > 0$, every Markov strategy has an $\varepsilon$-best response, which is itself a Markov strategy.

We introduce the following notation. For every pair of strategies $(\sigma_1, \sigma_2)$ and every history $h = \langle \alpha^*, \{t_\alpha\}_{\alpha \leq \alpha^*}, \iota \rangle$, denote by $U(\sigma_1, \sigma_2; h)$ the expected payoff of Player 1 under $(\sigma_1, \sigma_2)$ in the subgame that starts at $h$. The quantities $(U(\sigma_1, \sigma_2; h))_h$ are related through the following recursive equation:

$$U(\sigma_1, \sigma_2; h) = \int_{s_1, s_2 \in (t_{\alpha^*}, \infty]} \left[ \left( 1_{s_1 < s_2} + \frac{1}{2} \cdot 1_{s_1 = s_2} \right) e^{-r(s_1 - t_{\alpha^*})} \right. \quad (1)$$
$$\cdot \left( -c + (1 - e^{-\lambda(s_1 - t_{\alpha^*})}) v_1 + e^{-\lambda(s_1 - t_{\alpha^*})} U(\sigma_1, \sigma_2; h') \right)$$
$$+ \left( 1_{s_1 > s_2} + \frac{1}{2} \cdot 1_{s_1 = s_2} \right) e^{-r(s_1 - t_{\alpha^*})}$$
$$\left. \cdot \left( (1 - e^{-\lambda(s_2 - t_{\alpha^*})}) v_2 + e^{-\lambda(s_2 - t_{\alpha^*})} U(\sigma_1, \sigma_2; h') \right) \right] \sigma_1(h)(ds_1) \sigma_2(h)(ds_2),$$

where $h'$ is the history after the current one, so it is created by adding to $h$ either $(t_{\alpha^*+1} = s_1, \iota_a(\alpha^*+1) = \iota_i(\alpha^*+1) = \{1\})$ (if $s_1 < s_2$), or $(t_{\alpha^*+1} = s_2, \iota_a(\alpha^*+1) = \iota_i(\alpha^*+1) = \{2\})$ (if $s_1 > s_2$), or $(t_{\alpha^*+1} = s_1, \iota_a(\alpha^*+1) = \{1,2\}, \iota_i(\alpha^*+1) = \{1\})$ (if $s_1 = s_2 < \infty$ and Player 1 actually inspected), and $(t_{\alpha^*+1} = s_2, \iota_a(\alpha^*+1) = \{1,2\}, \iota_i(\alpha^*+1) = \{2\})$ (if $s_2 = s_1 < \infty$ and Player 2 actually inspected).

Denote

$$\widetilde{U}(\sigma_1, \sigma_2, h) := \int_{s_1, s_2 \in (t_{\alpha^*}, \infty]} \left( \left( 1_{s_1 < s_2} + \frac{1}{2} \cdot 1_{s_1 = s_2} \right) e^{-r(s_1 - t_{\alpha^*})} \left( -c + (1 - e^{-\lambda(s_1 - t_{\alpha^*})}) v_1 \right) \right.$$
$$\left. + \left( 1_{s_1 > s_2} + \frac{1}{2} \cdot 1_{s_1 = s_2} \right) e^{-r(s_1 - t_{\alpha^*})} \left( (1 - e^{-\lambda(s_2 - t_{\alpha^*})}) v_2 \right) \right) \sigma_1(h)(ds_1) \sigma_2(h)(ds_2),$$

the unconditional expected payoff of Player 1 due to the first inspection done after $t_{\alpha^*}$ (normalized to time $t_{\alpha^*}$), and by

$$\widetilde{P}(\sigma_1, \sigma_2, h) := \int_{s_1, s_2 \in (t_{\alpha^*}, \infty]} e^{-r(\min\{s_1, s_2\} - t_{\alpha^*})} \sigma_1(h)(ds_1) \sigma_2(h)(ds_2),$$

the expected discounted time (normalized to time $t_{\alpha^*}$) until the next inspection. Finally, define

$$\Lambda(\sigma_1, \sigma_2, h) := \widetilde{U}(\sigma_1, \sigma_2, h) / \widetilde{P}(\sigma_1, \sigma_2, h),$$

to be the normalized expected payoff to Player 1 until the next inspection. When $\sigma_1$ and $\sigma_2$ are Markov, $\Lambda(\sigma_1, \sigma_2, h)$ is independent of $h$, and coincides with Player 1's expected payoff under $(\sigma_1, \sigma_2)$.

Player 1's expected payoff under $(\sigma_1, \sigma_2)$ is a convex combination of $(\Lambda(\sigma_1, \sigma_2, h))_h$, where the weight of $\Lambda(\sigma_1, \sigma_2, h)$ is given by the unconditional probability that the first inspection done after history $h$ is successful.

Fix now an $\varepsilon > 0$ and a Markov strategy $\sigma_2$ of Player 2, and let $\sigma_1$ be an ($\varepsilon/2$)-best response of Player 1 to $\sigma_2$. We will show that Player 1 has an $\varepsilon$-best response which is Markovian, thereby completing the proof of the proposition.

Denote $u := \sup_h \Lambda(\sigma_1, \sigma_2, h)$, where the supremum is over all histories $h$. Since the expected payoff under $(\sigma_1, \sigma_2)$ is a convex combination of $(\Lambda(\sigma_1, \sigma_2, h))_h$, it follows that this expected payoff is at most $u$.

Let $h_0$ be a history such that $\Lambda(\sigma_1, \sigma_2, h_0) \geq u - \frac{\varepsilon}{2}$. Finally, let $\sigma_1'$ be the Markov strategy that is defined by $\sigma_1(h_0)$: after each inspection, the distribution of the next inspection time of Player 1 is according to $\sigma_1(h_0)$, shifted to the time of the last inspection.

Since $\sigma_1'$ and $\sigma_2$ are Markov strategies, the expected payoff under $(\sigma_1', \sigma_2)$ is $\Lambda(\sigma_1', \sigma_2, h)$ (and this quantity is independent of $h$). By the choice of $h_0$, this expected payoff is at least $u - \varepsilon/2$, and hence at least the expected payoff under $(\sigma_1, \sigma_2)$ minus $\varepsilon$.